\documentclass[a4paper,fleqn,usenatbib]{mnras}

\usepackage{newtxtext,newtxmath}

\usepackage[T1]{fontenc}
\usepackage{ae,aecompl}

\usepackage{graphicx}	
\usepackage{amsmath}	
\usepackage{amssymb}	
\usepackage{times}
\usepackage{graphics, subfigure}
\usepackage{epsfig}
\usepackage{multirow}
\usepackage{ulem}
\usepackage{pdfpages}

\def\simlt{\mathrel{\hbox{\rlap{\hbox{\lower4pt\hbox{$\sim$}}}\hbox{$<$}}}}
\def\simgt{\mathrel{\hbox{\rlap{\hbox{\lower4pt\hbox{$\sim$}}}\hbox{$>$}}}}

\newcommand{\mysim}{\mathord{\sim}}

\title[Late-time light curves]{
Constraints on the density distribution of type Ia supernovae ejecta inferred from late-time light-curve flattening
}
\author[D. Kushnir and E. Waxman]{
Doron Kushnir$^{1}$\thanks{E-mail: doron.kushnir@weizmann.ac.il} and Eli Waxman$^{1}$
\\
$^{1}$Dept. of Particle Phys. \& Astrophys., Weizmann Institute of
Science, Rehovot 76100, Israel
}

\date{Accepted XXX. Received YYY; in original form ZZZ}

\pubyear{2019}

\begin{document}
\label{firstpage}
\pagerange{\pageref{firstpage}--\pageref{lastpage}}
\maketitle

\begin{abstract}
The finite time, $\tau_{\rm dep}$, over which positrons from $\beta^{+}$ decays of $^{56}$Co deposit energy in type Ia supernovae ejecta lead, in case the positrons are trapped, to a slower decay of the bolometric luminosity compared to an exponential decline. Significant light-curve flattening is obtained when the ejecta density drops below the value for which $\tau_{\rm dep}$ equals the $^{56}$Co life-time. We provide a simple method to accurately describe this "delayed deposition" effect, which is straightforward to use for analysis of observed light curves. We find that the ejecta heating is dominated by delayed deposition typically from 600 to 1200~day, and only later by longer lived isotopes $^{57}$Co and $^{55}$Fe decay (assuming solar abundance). For the relatively narrow $^{56}$Ni velocity distributions of commonly studied explosion models, the modification of the light curve depends mainly on the $^{56}$Ni mass-weighted average density, $\langle \rho \rangle t^{3}$. Accurate late-time bolometric light curves, which may be obtained with JWST far-infrared (far-IR) measurements, will thus enable to discriminate between explosion models by determining $\langle \rho \rangle t^3$ (and the $^{57}$Co and $^{55}$Fe abundances). The flattening of light curves inferred from recent observations, which is uncertain due to the lack of far-IR data, is readily explained by delayed deposition in models with $\langle \rho\rangle t^{3} \approx 0.2\,M_{\odot}\,(10^{4}\, \textrm{km}\,\textrm{s}^{-1})^{-3}$, and does not imply supersolar $^{57}$Co and $^{55}$Fe abundances. 
\end{abstract}




\section{Introduction}
\label{sec:Introduction}

It is widely accepted that the light curves of both Type Ia supernovae (Ia SNe) and, at least at late times, core collapse supernovae (CCSNe, including type Ib/c), are powered by the decay of radionuclides synthesized in the explosion. The most important power source is the decay-chain \citep{Pankey1962,Colgate1969}
\begin{eqnarray}\label{eq:56 chain}
^{56}\textrm{Ni}\xrightarrow{t_{1/2}=6.07\,\textrm{d}}{}^{56}\textrm{Co}\xrightarrow{t_{1/2}=77.2\,\textrm{d}}{}^{56}\textrm{Fe}.
\end{eqnarray}
Part of the decay energy of this chain is released in the form of $\sim1\,\textrm{MeV}$ $\gamma$-rays, while the second stage of the chain includes also a $\beta^{+}$ decay channel in which $\sim1\,\textrm{MeV}$ positrons are released. These positrons lose their kinetic energy and heat the plasma mainly by ionization, and then release their rest-mass energy by annihilation. The contribution of this chain to the energy deposition rate is given at late times by \citep{Swartz1995,Jeffery1999}
\begin{eqnarray}\label{eq:56 deposition}
Q_{56}\approx\frac{M_{\rm{Co}56}}{M_{\odot}}e^{-t/111.4\,\textrm{d}}\left[138\left(\frac{t_0}{t}\right)^{2}+4.64\right]10^{41}\,\textrm{erg}\,\textrm{s}^{-1},
\end{eqnarray}
where $M_{\rm{Co}56}$ is the mass of $^{56}$Co and all its radioactive parents at the time of explosion, $t$ is the time since explosion and $t_0$ is the time at which the column density of the ejecta is comparable to the mean free path of $\gamma$-rays produced in the decay \citep[$t_0\approx35\,\textrm{day}$ for Ia SNe,][]{Wygoda2019}. The first term in the square brackets of Equation~\eqref{eq:56 deposition} represents the $\gamma$-ray energy deposition, valid at late times, $t>t_0$, when the probability of a $\gamma$-ray to experience a single (multiple) Compton collision before escaping is small (negligible). The second term represents the positron energy deposition, under the approximation that it is instantaneous. For times $t\gg t_0$, the main heating source of the ejecta is the kinetic energy loss of the positrons \citep{Arnett1979,Axelrod1980}.

Additional heating is provided by the decay-chains \citep{Seitenzahl2009}
\begin{eqnarray}\label{eq:57,55 chains}
&&^{57}\textrm{Ni}\xrightarrow{t_{1/2}=35.6\,\textrm{h}}{}^{57}\textrm{Co}\xrightarrow{t_{1/2}=271.2\,\textrm{d}}{}^{57}\textrm{Fe},\nonumber\\
&&^{55}\textrm{Co}\xrightarrow{t_{1/2}=17.53\,\textrm{h}}{}^{55}\textrm{Fe}\xrightarrow{t_{1/2}=999.7\,\textrm{d}}{}^{55}\textrm{Mn}.
\end{eqnarray}
While the abundance of these isotopes is low, their contribution may be significant at times longer than the $^{56}$Co life time, due to their longer life times. Although positrons are not emitted in the second stages of these chains, low-energy internal conversion (IC) and Auger electrons are emitted, along with their (less energetic) associated X-ray cascades. The contribution of these chains to the bolometric luminosity, assuming instantaneous energy deposition by the emitted particles, is
\begin{eqnarray}\label{eq:57,55 deposition}
Q_{57}&\approx&\frac{M_{\rm{Co}57}}{M_{\odot}}e^{-t/391.3\,\textrm{d}}2.22\times10^{40}\,\textrm{erg}\,\textrm{s}^{-1},\nonumber\\
Q_{55}&\approx&\frac{M_{\rm{Fe}55}}{M_{\odot}}e^{-t/1442.2\,\textrm{d}}1.52\times10^{39}\,\textrm{erg}\,\textrm{s}^{-1},
\end{eqnarray}
where $M_{\rm{Co}57}(M_{\rm{Fe}55})$ is the mass of $^{57}$Co($^{55}$Fe) and all its radioactive parents at the time of explosion. 

At the late times of interest here, the radioactive heating of the ejecta is dominated by the energy deposition of electrons and positrons, and the bolometric luminosity is well approximated by the energy deposition rate (see discussion at \S~\ref{sec:LQ}). As a result, the observed light curve depends on the electron/positron transport, which in turn depends on the magnetic field within the expanding ejecta. In the presence of a sufficiently strong tangled magnetic field, the electrons/positrons are "fully trapped," that is, confined to the fluid element in which they were released. In the absence of such a field, electrons/positrons may escape the ejecta.

The dependence of the light curves on the transport of electrons/positrons was studied in the context of Ia SNe using Monte-Carlo simulations for some specific ejecta and magnetic field configurations \citep{Colgate1980,Milne1999,Milne2001}. A few approximate calculation methods were suggested as well \citep{Chan1993,Cappellaro1997,Ruiz1998}. \citet{Milne1999} found that for fully trapped positrons the finite positron energy deposition time, $\tau_{\rm dep}$, combined with the exponential decline in the number of newly created positrons, leads to a flattening of the light curve, as the positrons' kinetic energy is stored and contributes to ejecta heating at later times \citep[\textit{the delayed deposition effect}, the effect is also briefly discussed by][]{Axelrod1980}. On the other hand, if positrons escape the ejecta, then the kinetic energy loss leads to a steepening of the light curve. The results of Monte-Carlo simulations were compared to optical light-curve observations in a few cases \citep{Colgate1980,Milne1999,Milne2001,Milne2003}, with the conclusion that there is evidence for a significant escape of positrons. However, with the additional measurements of redder bands \citep{Lair2006a,Lair2006b} and especially infrared (IR) \citep{Sollerman2004,Spyromilio2004}, it was discovered that the bolometric light curves do not show a steepening as would be expected in the case of positron escape, but rather an increase with time of the fraction of light emitted in the redder bands \citep[see also][]{Bryngelson2012,Graur2020}. With a lack of evidence for positron escape, we consider in what follows only the fully trapped positrons scenario.

In the last few years, a few Ia SNe were observed to late times with a more complete spectral coverage, and were found to exhibit a late-time flattening of the light curve \citep{Graur2016,Dimitriadis2017,Kerzendorf2017,Shappee2017,Yang2018,Graur2018,Graur2018b,Graur2019,Li2019}. Assuming that the observed flattening is due to heating by $^{57}$Co and $^{55}$Fe decays, and ignoring delayed deposition, abundances of $^{57}$Co and $^{55}$Fe were derived, typically assuming fully trapped positrons. As we show here, delayed deposition by trapped positrons competes with, and may dominate, plasma heating by $^{57}$Co and $^{55}$Fe decays. Thus, ignoring the contribution of the delayed deposition effect to the flattening of the light curves would lead to a systematic overestimate of the abundances.

In this paper we provide a simple and accurate method to calculate the delayed deposition effect, bypassing the use of Monte-Carlo simulations that makes the scan of a large model parameter space difficult. The method was developed in the context of "kilonovae", i.e. for calculating light curves produced by the ejecta of neutron-star mergers \citep{Waxman2018,Waxman2019}. Given the large uncertainties regarding the structure and composition of neutron-star mergers' ejecta, results were derived for general simple distributions of the electrons/positrons' energy and of the ejecta velocity. Here we apply this method for the study of positrons from $\beta^{+}$ decays of $^{56}$Co, that are fully trapped in an iron Ia SN ejecta. This enables us to derive exact results for ejecta velocity distributions of commonly studied explosion models.

Two comments are in place here regarding approximations adopted in our analysis. First, we calculate the energy loss of the positrons/electrons assuming that they are embedded in a plasma dominated by Iron group elements (for which the energy loss is well approximated by that of an Iron plasma). This is a valid approximation since most of the $^{56}$Ni mass produced in common Ia SN models is contained in a plasma dominated by Iron group elements, and since the energy loss depends only weakly on the plasma composition (see Section~\ref{sec:e_dep}). Expanding our analysis method to include energy loss in a plasma of different composition is straight forward. Secondly, we calculate energy losses assuming a time-independent low ionization, with a number of free electrons per atom $\chi_e\sim1$. This is a valid approximation since, as we show in Section~\ref{sec:non-instantaneous}, the results are not sensitive to the exact value of $\chi_e$, and since at the relevant times the plasma ionization cannot be significantly larger. Here too, expanding our analysis method to include a more accurate description of $\chi_e$ is straight forward.

The structure of the paper is as follows. In Section~\ref{sec:non-instantaneous}, we describe our calculation method, derive the late-time bolometric light curves and discuss the effect of delayed deposition. We consider radioactive heating by the decay of $^{56}$Co, $^{57}$Co, and $^{55}$Fe, and assume that the bolometric luminosity equals the heating rate, as appropriate at the late times of interest. We focus on ejecta velocity distributions of commonly studied Ia SNe explosion models \citep["Chandra", "sub-Chandra" and direct-collision models, see][for a review]{Maoz2014}, but address also the modifications to the light curves that will arise in the presence of wider velocity distributions. In Section~\ref{sec:SN 2011fe}, we demonstrate how ejecta properties may be inferred from observations by analyzing the late-time light curve of SN 2011fe using our method. Our conclusions are summarized and discussed in Section~\ref{sec:discussion}, where we also comment on the differences in the role played by delayed deposition in the different types of ejecta produced in Ia SNe, CCSNe, and kilonovae.


\section{A simple and accurate method to calculate the delayed deposition effect}
\label{sec:non-instantaneous}

We consider the bolometric luminosity produced by a radioactively heated expanding ejecta. We are interested in the evolution at late times, at which the contribution of $\gamma$-rays to the energy deposition is small and given by the first term in the square brackets of Equation~(\ref{eq:56 deposition}). The equations describing the evolution of the kinetic energy $E$ of electrons/positrons released in radioactive decays are given in Section~\ref{sec:e energy}. The equations describing the heating of the ejecta by confined electrons/positrons are given in Section~\ref{sec:e_dep}. In Section~\ref{sec:late LC}, we discuss the effects of delayed deposition on the late-time light curves. In our analysis, we approximate the bolometric luminosity as given by the energy deposition rate, $L=Q$. We show in Section~\ref{sec:LQ} that this is a valid approximation for the time range in which we are interested.

\subsection{The evolution of electron/positron energy}
\label{sec:e energy}

We assume that electrons/positrons produced in radioactive decays are "fully trapped", that is, confined to the fluid elements in which they were released. The trapped particles lose energy by ionization, free electron scattering, and bremsstrahlung, that contribute to plasma heating, and by adiabatic losses due to plasma expansion, that do not lead to heating but rather to the acceleration of the ejecta. The kinetic energy $E$ of an electron/positron evolves according to
\begin{eqnarray}\label{eq:E(t)}
\frac{dE}{dt}=-v_e\left(\frac{p_e}{t}+\rho \frac{dE}{dX}\right).
\end{eqnarray}
Here, $v_e$ and $p_e$ are the electron/positron's velocity and momentum, $\rho$ is the plasma density, and $dX=\rho v_e dt$ is the column density traversed by the electron/positron in the time interval $dt$. The second term describes the energy losses due to ionization, scattering, and bremsstrahlung, while the first term describes adiabatic losses assuming that (in the absence of other losses) adiabatic expansion leads to $p_e\propto 1/t$. The velocity, $\bf{v}$, of each fluid element in the ejecta is constant, and its density decreases as $\rho\propto t^{-3}$. For spherical symmetry, the density is related to the mass--velocity distribution of the ejecta by
\begin{equation}\label{eq:rho}
  4\pi t^3\rho(v,t)=v^{-2}dm/dv,
\end{equation}
where $m(v)$ is the mass of the ejecta with velocity $<v$.

The ionization, scattering and, bremsstrahlung loss processes are briefly described below.

\begin{enumerate}

\item \textbf{Ionization} -- The energy loss per unit column density (grammage) traversed by the electron/positron, $dX=\rho dx$, is given by \citep[e.g.,][]{Longair1992}
\begin{eqnarray}\label{eq:ionization losses}
\left(\frac{dE}{dX}\right)_{\rm{ion}}&=&\frac{4\pi e^4}{m_e m_p v_e^2}\frac{Z}{A}\left[\ln\left(\frac{\gamma_e^2m_ev_e^2}{\bar{I}}\right)-\frac{1}{2}\ln\left(1+\gamma_e\right)\right.\nonumber\\
 &-&\left. \left(\frac{2\gamma_e+\gamma_e^2-1}{2\gamma_e^2}\right)\ln2+\frac{1}{2\gamma_e^2}+\frac{1}{16}\left(1-\frac{1}{\gamma_e}\right)^2\right],
\end{eqnarray}
where $\gamma_e$ is the Lorentz factor of the electron/positron, $Z$ and $A$ are the atomic and mass numbers of the plasma nuclei, and $\bar{I}$ is the effective average ionization energy of the atoms, which is empirically determined and approximately given by $\bar{I}=10Z\,\textrm{eV}$ for heavy nuclei \citep[e.g.,][]{Mukherji1975}. In general, for ionized plasma one should use the number of bound electrons and the effective $\bar{I}$ for the ionized state. However, since we consider only low-ionization plasma, and since the dependence on $\bar{I}$ is logarithmic, these corrections are small.

\item \textbf{Plasma losses} -- Electrons/positrons also lose energy by scattering free plasma electrons, with a rate given by \citep{Solodov2008}
\begin{eqnarray}\label{eq:plasma losses}
\left(\frac{dE}{dX}\right)_{\rm{plasma}}&=&\frac{4\pi e^4}{m_e m_p v_e^2}\frac{\chi_e}{A}\left\{\ln\left[\left(\frac{E}{\hbar \omega_p}\right)^2\frac{\gamma_e+1}{2\gamma_e^2}\right]+1\right.\nonumber\\
 &+&\left. \frac{1}{8}\left(\frac{\gamma_e-1}{\gamma_e}\right)^2-\left(\frac{2\gamma_e-1}{\gamma_e^2}\right)\ln 2\right\},
\end{eqnarray}
where $\chi_e$ is the number of free electrons per atom, $\omega_p=(4\pi e^2 n_e /m_e)^{1/2}$ is the plasma frequency, and $n_e=\chi_e\rho/(A m_p)$ is the number density of free electrons.

\item \textbf{Bremsstrahlung} -- At highly relativistic energy, bremsstrahlung losses dominate, with a rate given by \citep[e.g.,][]{Longair1992}
\begin{eqnarray}\label{eq:brem}
\left(\frac{dE}{dX}\right)_{\rm{Brem}}=\frac{4 e^4}{m_e m_p c v_e}\frac{Z^2e^2}{A\hbar c}\frac{E}{m_e c^2}\left[\ln\left(\frac{183}{Z^{1/3}}\right)+\frac{1}{8}\right].
\end{eqnarray}

\end{enumerate}

\begin{figure}
\includegraphics[width=0.48\textwidth]{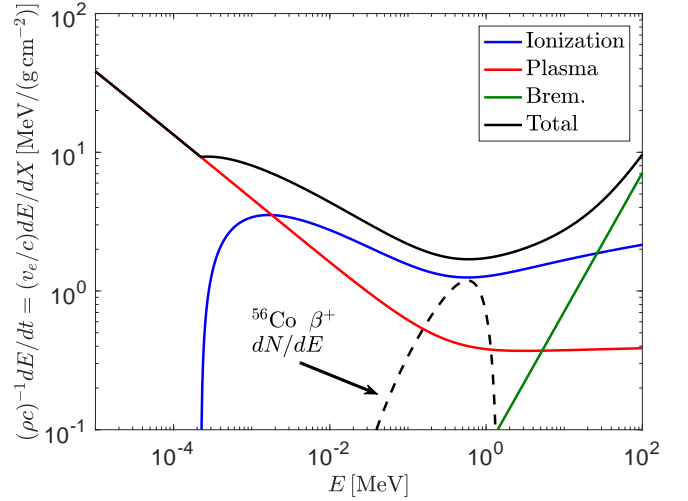}
\caption{The energy loss rate of electrons/positrons propagating in a singly ionized, $\chi_e=1$, iron plasma ($Z=26$, $A=56$), as a function of their kinetic energy. Blue: ionization losses, Equation~\eqref{eq:ionization losses}; red: plasma losses; Equation~\eqref{eq:plasma losses}; green: bremsstrahlung losses, Equation~\eqref{eq:brem}; black: the total loss rate.  Also shown (dashed black line) is the energy spectrum of positrons released in $\beta^{+}$ decay of $^{56}$Co, Equation~\eqref{eq:dEdN}. The energy loss rate for a typical $\beta^{+}$ positron is dominated by ionization losses with some tens of percent contribution from plasma losses, leading to a total energy loss rate of $(v_e/c)dE/dX\approx 2\,\textrm{MeV}/(\textrm{g}\,\textrm{cm}^{-2})$.
\label{fig:Stopping}}
\end{figure}

The energy-dependent energy loss rate of electrons/positrons propagating in a singly ionized, $\chi_e=1$, iron plasma ($Z=26$, $A=56$) is shown in Figure~\ref{fig:Stopping}. Also shown is the energy spectrum of positrons released in $\beta^{+}$ decay of $^{56}$Co, given by \citep[e.g.,][]{Nadyozhin1994}
\begin{eqnarray}\label{eq:dEdN}
\frac{dN}{dE}\propto\left(E+m_ec^2\right)^2\left(1.4589\,\textrm{MeV}-E\right)^2G(E),
\end{eqnarray}
where the function $G(E)$ is tabulated in \citet{Rose1955}. As can be seen, the energy loss rate for a typical $\beta^{+}$ positron is dominated by ionization losses with some tens of percent contribution from plasma losses, leading to a total energy loss rate of $(v_e/c)dE/dX\approx 2\,\textrm{MeV}/(\textrm{g}\,\textrm{cm}^{-2})$. For lower energy (Auger and IC) electrons, the energy loss rate is higher.

We next define an effective electron/positron 'opacity'
\begin{eqnarray}\label{eq:opacity}
\kappa_e\equiv \left[\left(dE/dX\right)/E\right]_{E=1\,\textrm{MeV}}\approx2\,\textrm{cm}^2\,\textrm{g}^{-1}.
\end{eqnarray}
The energy loss rate can be written as
\begin{eqnarray}\label{eq:E(t) loss}
\frac{1}{\rho c}\frac{dE}{dt}=\frac{v_e}{c}\frac{dE}{dX}=1\kappa_e g(E) {\rm MeV},
\end{eqnarray}
where $g(E)\approx1$ is a weak function of the energy in the energy regime relevant for $^{56}$Co decay \citep{Waxman2019}.

\subsection{Energy deposition by electrons/positrons}
\label{sec:e_dep}

In this section, we derive the equations governing the energy deposition by electrons/positrons \citep[following][]{Waxman2019}, in a single fluid element with a density $\rho(t)$. The deposition in a model ejecta with a range of densities may be obtained straightforwardly by integrating over the independent contributions of the different fluid elements.

Let us consider first electrons/positrons produced with a single energy $E_{i}$, and denote by $E(t_i,t)$ the time-dependent energy of a positron produced at time $t_i$. The heating rate of the plasma at time $t$ is then given by
\begin{eqnarray}\label{eq:plasma heating 1}
Q_d(t,E_i)=\rho(t) \int dE \frac{\partial n(E,t)}{\partial E}v_e\left[\left(\frac{dE}{dX}\right)_{\rm{ion}}+\left(\frac{dE}{dX}\right)_{\rm{plasma}}\right],
\end{eqnarray}
where $dn/dE$ is the number of electrons/positrons per unit energy. We neglect the bremsstrahlung contribution, which is negligible for the relevant electron/positron energies (see Figure~\ref{fig:Stopping}).

The differential electron/positron number is given by
\begin{eqnarray}\label{eq:diff n}
\frac{\partial n(E,t)}{\partial E}=\dot{n}\left[t_i(E,t)\right]\frac{\partial t_i}{\partial E},
\end{eqnarray}
where $\dot{n}$ is the production rate of electrons/positrons by radioactive decay, and $t_i(E,t)$ is the time at which a positron should be produced in order to have an energy $E$ at time $t$. Using this relation, we may write Equation~\eqref{eq:plasma heating 1} as
\begin{eqnarray}\label{eq:plasma heating 2}
Q_d(t,E_i)=\rho(t) \int^{t} dt_{i} \dot{n}(t_i)\left\{v_e\left[\left(\frac{dE}{dX}\right)_{\rm{ion}}+\right.\right.\nonumber\\
\left.\left.\left(\frac{dE}{dX}\right)_{\rm{plasma}}\right]\right\}_{E(t_i,t)}.
\end{eqnarray}

The energy distribution of the positrons produced in the $\beta$ decay is taken into account by averaging $Q_d(t,E_i)$ with weights proportional to the positron energy distribution, $dN/dE$, yielding $Q_d(t)$. For $^{56}$Co decay, the positron energy distribution is given by Equation~\eqref{eq:dEdN}. We provide a Matlab code to calculate numerically $Q_d(t)$\footnote{Available through https://www.dropbox.com/sh/k4r9dyuqwhgbvdr/AACLNZ-m-{}-{}-x8h9QgtjdLIxfa?dl=0}.

\subsection{The effects of delayed deposition on the late-time light curves}
\label{sec:late LC}

We now turn to calculate the effects of delayed deposition on the late-time light curves. The energy deposition by positrons produced in $^{56}$Co $\beta^{+}$ decays is given by the equations derived in the preceding sub-section. For the energy deposition by $^{57}$Co and $^{55}$Fe we approximate the energy deposition by electrons as instantaneous, and use Equation~(\ref{eq:57,55 deposition}). This approximation is valid since the energy of the electrons is lower, with energy loss rate significantly higher, compared to that of $^{56}$Co positrons, and since the life times of $^{57}$Co and $^{55}$Fe are significantly longer than that of $^{56}$Co. The combination of the shorter loss time and longer decay time implies that, at the relevant times, the energy deposition time is small compared to the $^{57}$Co and $^{55}$Fe life times, and the effect of delayed energy deposition is small.

There are two characteristic times, at which the light-curve behaviour changes qualitatively due to the delayed deposition effect. As the density of the ejecta decreases, the positrons lose energy to ionization and scattering at a slower rate. When the density is sufficiently low, so that the energy deposition time $\tau_{\rm dep}$ exceeds the $^{56}$Co half life time, $t_{1/2}$, the effect of delayed deposition becomes important. Estimating $\tau_{\rm dep}$ using Equation~(\ref{eq:E(t) loss})
\begin{eqnarray}\label{eq:tau_dep}
  \tau_{\rm dep}=E_{\rm MeV}/(\rho c \kappa_e g(E_i)),
\end{eqnarray}
where $E_i=1E_{\rm MeV}$~MeV is the average initial positron energy, we define $t_{\tau}$ as the time at which $\tau_{\rm dep}=t_{1/2}$,
\begin{eqnarray}\label{eq:t tau}
t_{\tau}&=&\left(t_{1/2}\langle\rho\rangle t^3 c \kappa_e g(E_i) E^{-1}_{\rm MeV}\right)^{1/3}\nonumber\\
&\approx&630\left(\frac{\langle \rho t^3 \rangle_{\rm{SND}}}{0.2}E^{-1}_{\rm MeV}\right)^{1/3}\,\textrm{day}.
\end{eqnarray}
Here, $\langle\rho\rangle$ is the $^{56}$Ni (and its daughter nuclei) mass-averaged density, and we measure $\langle\rho\rangle t^3$ in units characteristic of the density of Ia SN ejecta
\begin{eqnarray}\label{eq:SND}
\langle \rho t^3 \rangle_{\rm{SND}}\equiv\frac{\langle \rho \rangle t^3}{1M_{\odot}/\left(10^{4}\,\textrm{km}\,\textrm{s}^{-1}\right)^{3}} = \frac{\langle v^{-2}dm/dv \rangle}{4\pi M_{\odot}/\left(10^{4}\,\textrm{km}\,\textrm{s}^{-1}\right)^{3}}.
\end{eqnarray}
Here, the last equality holds for spherically symmetric ejecta. At still later times, $\tau_{\rm dep}$ exceeds the expansion time, and adiabatic energy losses become dominant. We define $t_{\varepsilon}$ as the time at which $\tau_{\rm dep}=t$
\begin{eqnarray}\label{eq:t eps}
t_{\varepsilon}&=&\left(\langle \rho \rangle t^3 c \kappa_e g(E_i) E^{-1}_{\rm MeV}\right)^{1/2}\nonumber\\
&\approx&1790\left(\frac{\langle \rho t^3 \rangle_{\rm{SND}}}{0.2}E^{-1}_{\rm MeV}\right)^{1/2}\,\textrm{day}.
\end{eqnarray}
As we show below, for the relatively narrow $^{56}$Ni velocity distributions of commonly studied Ia SN models, the range of densities over which radioactive decays take place is limited, such that the light-curve modifications are largely determined by the $^{56}$Ni mass-weighted average density, $\langle \rho \rangle t^3$.

\begin{figure}
\includegraphics[width=0.53\textwidth]{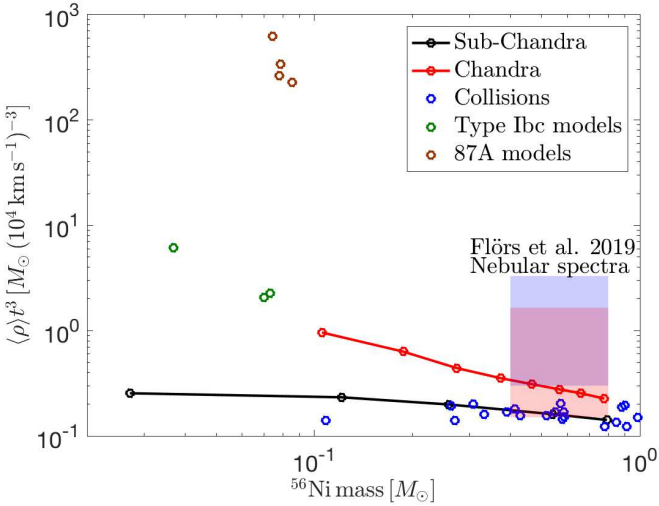}
\caption{$\langle \rho \rangle t^3$ values for various supernova explosion models, as a function of the synthesized $^{56}$Ni mass. Black: central detonation of sub-Chandra models with solar metallicity \citep[similar to the models of][as will be reported in a subsequent publication]{Shen2018}. The masses of the exploding CO WDs are $0.8,0.85,0.9,1,1.1\,M_{\odot}$ (left to right). Red: Chandra models \citep{Dessart2014}. Blue: direct-collision models \citep{Kushnir2013}. Green: Type Ib/c models \citep{Dessart2016,Yoon2019}. Brown: SN 1987A models \citep{Blinnikov2000,Utrobin2005,Sukhbold2016,Dessart2019}. Also shown are the estimates of \citet{Flors2019}, based on nebular-spectra analysis of high-luminosity Ia SNe (blue/red shaded area for singly/doubly ionized plasma, where the overlapping region is in purple).
\label{fig:rhot3}}
\end{figure}

Figure~\ref{fig:rhot3} shows the $\langle \rho \rangle t^3$ values of various  supernovae models. For high-luminosity Ia SNe, both sub-Chandra models (black line) and direct-collision models (blue circles) predict similar low values, $\langle \rho t^3 \rangle_{\rm{SND}}\approx0.15$, while the Chandra models (red line) predict somewhat higher values, $\langle \rho t^3 \rangle_{\rm{SND}}\approx0.25$. The difference between the models increases for low-luminosity Ia SNe. We also show the estimates of \citet{Flors2019}, based on a nebular-spectra analysis of high-luminosity Ia SNe (blue/red shaded areas for singly/doubly ionized plasma, where the overlapping region is in purple). While the uncertainty is quite substantial, the models predict values that are at the low end of these estimates. For CCSNe, much larger values are expected, $\langle \rho t^3 \rangle_{\rm{SND}}\approx1\,{\rm to\,} 10$ for Type Ib/c (green circles) and $\langle \rho t^3 \rangle_{\rm{SND}}\approx100\,{\rm to\,}1000$ for SN 1987A models (brown circles). This implies that the effects of delayed deposition would be significant for CCSNe at much later times compared to Ia SNe.

The delayed deposition of positron energy at $t>t_\tau$ leads to a flattening of the light curve, that is, to an enhancement of the emission above an exponential decline following the decline of the $^{56}$Co decay rate, that would be obtained for instantaneous energy deposition, $Q_{ins}(t)$. To see this, note that when $t>t_\tau$ we have $\tau_{\rm dep}(t)>t_{1/2}$, and positrons produced at time $t$ deposit their energy over a time longer than the exponential decay time of the $^{56}$Co decay rate. Thus, at time $t+\tau_{\rm dep}(t)$ the population of positrons that heat the plasma is dominated by positrons produced at time $t$, the number of which is larger than that of positrons produced at $t+\tau_{\rm dep}(t)$ by a factor $\exp{[\tau_{\rm dep}(t)/t_{1/2}]}$. While the kinetic energy $E$ of the positrons produced earlier is lower, they dominate the heating since their number is larger and the ionization/plasma loss rate is nearly independent of $E$ (see Figure~\ref{fig:deposition}). The enhanced late-time deposition and luminosity, compared to that produced by instantaneous deposition, is produced by the energy that is "stored" in the positrons that did not lose all their energy instantaneously at earlier time. Due to the fast exponential decrease in the decay rate, the small fractional reduction of the deposition (and luminosity) at early time, obtained for a finite deposition time compared to instantaneous deposition, translates to a large fractional contribution to the deposition and luminosity at late time. The small fractional deficit at earlier times will be typically difficult to detect (see below).

In Figure~\ref{fig:to_t_demo} we show the time $\Delta t(E_i,t)$ over which a positron of initial energy $E_i$ produced at time $t$ loses all its energy, obtained by numerically solving Equation~\eqref{eq:E(t)} for a few initial energies and two values of the plasma density. Solving this equation, the positron reaches zero energy at a finite time. In reality, its energy loss is suppressed when it reaches the thermal plasma energy. Since the thermal energy is much smaller than $E_i$, the implied corrections to the deposited energy and to $\Delta t(E_i,t)$ are small. For the average energy of $^{56}$Co $\beta^{+}$ decay positrons, $0.63\,\textrm{MeV}$, we find that the time at which the energy loss time equals the $^{56}$Co half life time is $t_{\tau}\approx700(1200)\,\textrm{day}$ for $\rho t^3=0.2(1) M_\odot/(10^4{\rm km\,s^{-1}})^3$, in agreement with the estimate of Equation~(\ref{eq:t tau}). The modification of the light curve is significant at $t>t_\tau$. For $t_\tau$ smaller than $t_{57}$, the time  at which the instantaneous energy deposition rates by $^{57}$Co and $^{56}$Co are equal, delayed deposition dominates the contribution to the plasma heating by $^{57}$Co decay over a significant time.

\begin{figure}
\includegraphics[width=0.48\textwidth]{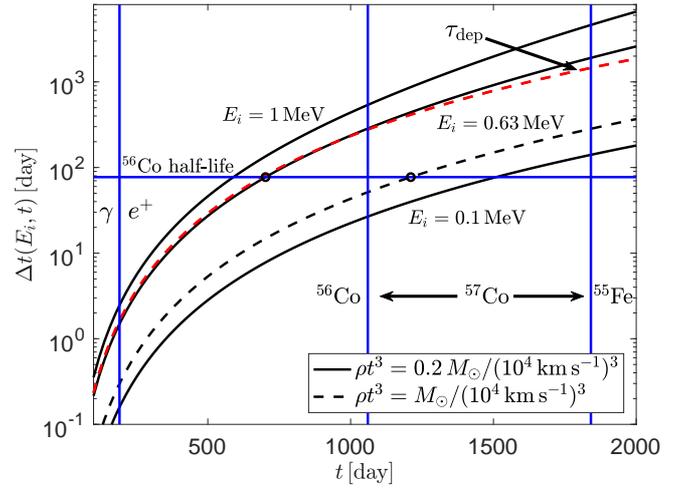}
\caption{The time $\Delta t(E_i,t)$ over which a positron of initial energy $E_i$ produced at time $t$ loses all its energy, obtained by numerically solving Equation~\eqref{eq:E(t)} for a few initial energies and two values of the plasma density, assuming singly ionized plasma (black curves) . The $^{56}$Co half-life $t_{1/2}\approx77.2\,\textrm{day}$ is indicated by the blue horizontal line. The red dashed line shows $\tau_{\rm dep}$, given by Equation~(\ref{eq:tau_dep}), for $E_i=0.63\,\textrm{MeV}$ and $\rho t^3=0.2M_\odot/(10^4{\rm km\,s^{-1}})^3$. $\Delta t(E_i,t)$ is larger than $\tau_{\rm dep}$ at late time since the density and loss rate decrease with time. The two rightmost blue vertical lines indicate the times at which the instantaneous energy deposition rate from $^{56}$Co equals the deposition rate by $^{57}$Co and $^{55}$Fe decays, $t_{57}$ and $t_{55}$ respectively, assuming solar metallicity \citep[mass ratios $^{57}\textrm{Fe}/^{56}\textrm{Fe}=0.023$ and $^{55}\textrm{Mn}/^{56}\textrm{Fe}=0.011$,][]{Lodders2019}. The leftmost blue vertical line indicates the time at which the $\gamma$-ray and the positron energy deposition rates are equal for Type Ia SNe ($t_e\approx5.45 t_0\approx190\,\textrm{day}$). The black circles show the times at which $\Delta t=t_{1/2}$ for $E_i=0.63\,\textrm{MeV}$, the average energy of $^{56}$Co $\beta^{+}$ decay positrons, $\approx700(1200)\,\textrm{day}$ for $\rho t^3=0.2(1) M_\odot/(10^4{\rm km\,s^{-1}})^3$, in agreement with the estimate of Equation~(\ref{eq:t tau}). For $t_\tau<t_{57}$, delayed deposition dominates the contribution to the plasma heating by $^{57}$Co decay over a significant time.
\label{fig:to_t_demo}}
\end{figure}

Figure~\ref{fig:deposition} shows the fractional modification of the heating rate and bolometric luminosity, compared to the heating rate that would be obtained under the instantaneous energy deposition approximation, for $ \rho t^3=0.161M_\odot/(10^4{\rm km\,s^{-1}})^3$, corresponding to the value of $\langle\rho\rangle t^3$ predicted by sub-Chandra detonation models with $M=1\,M_{\odot}$ and solar-metallicity WD. Results are shown for several values of $\chi_e$, the number of free electrons per ion, demonstrating that the sensitivity to the exact ionization level is small. At early times, the heating rate is suppressed by a few percent due to the finite energy loss time of the positrons. These positrons deposit their energy at later times, $t\gtrsim t_{\tau}$, leading to a large enhancement of the deposition rate at late time. By the time that the energy deposition by $^{57}$Co decays becomes significant, the enhancement due to delayed deposition is already large (order unity) and the $\gamma$-ray contribution to the plasma heating is small, making the identification of the delayed deposition effect straight forward.

\begin{figure}
\includegraphics[width=0.48\textwidth]{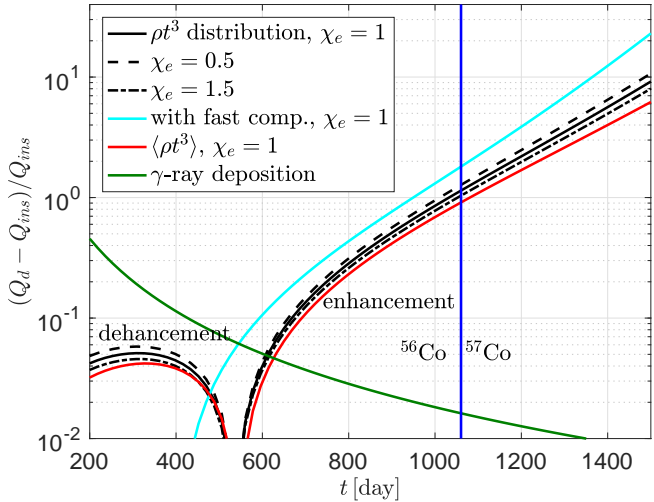}
\caption{The fractional difference between the numerically calculated delayed deposition rate, $Q_{d}(t)$, and the instantaneous deposition rate, $Q_{ins}(t)$ (calculated by ignoring the delayed deposition effect) of $^{56}$Co $\beta^{+}$ positrons. Black lines: Results for the full density distribution obtained in a sub-Chandra detonation model of a $1\,M_{\odot}$ solar-metallicity WD and several values of $\chi_e$, the number of free electrons per ion; Red: The results obtained for a single density corresponding to the $^{56}$Ni weighted average model density, $\langle \rho t^3 \rangle_{\rm{SND}}=0.161$; Cyan: The modification due to the addition of a high velocity component of $^{56}$Co, around $\rho t^3=10^{-3}M_\odot/(10^4{\rm km\,s^{-1}})^3$, with $10\%$ of the mass of the original model (for this model too, we use $Q_{ins}(t)$ excluding the contribution of the fast component, since $\gamma$-rays escape from it already at very early time, $t_0\approx13\,\textrm{day}$, so that the $^{56}$Ni mass deduced from the light curve would not include its contribution). At early times, there is a suppression of a few percent of the positron heating rate due to the finite time it takes positrons to deposit their energy. These positrons deposit their energy at later times, leading to a large enhancement of the deposition rate at $t\gtrsim t_{\tau}$. By the time that energy deposition from $^{57}$Co becomes important (blue vertical line, calculated for a solar metallicity mass ratio, $^{57}\textrm{Fe}/^{56}\textrm{Fe}=0.023$), the enhancement due to delayed deposition is already large (order unity) and the $\gamma$-ray contribution to the plasma heating (green line) is small, making the identification of the delayed deposition effect straight forward.
\label{fig:deposition}}
\end{figure}

Figure~\ref{fig:deposition} also shows a comparison of the deposition rate obtained using the full density distribution of the ejecta, as obtained in the sub-Chandra detonation model, and the deposition rate obtained using a single density corresponding to the average model density, $\langle\rho\rangle$. The difference is small. We found similar small differences between the results obtained using the average density and the full density distribution for direct-collision models and Chandra models. This implies that, for the commonly studied explosion models, using $\langle \rho  \rangle t^3$ instead of the full distribution provides a good approximation to the resulting deposition rates. Nevertheless, the strong dependence of $t_\tau$ on the velocity implies that the presence of a relatively small amount of $^{56}$Co within fluid elements with relatively high velocity, with correspondingly low values of $\rho t^{3}$, may have a large effect on the delayed deposition rate and bolometric luminosity, since the time at which delayed deposition would become important for such fast components would be significantly earlier than $t_\tau$ corresponding to the average density. For example, consider the double-detonation model of \citet{Polin2019} for a $1\,M_{\odot}$ WD with a helium shell of $0.08\,M_{\odot}$ (presented in their figure 1). In this model, a few percent of the total $^{56}$Ni is synthesized in the helium shell, with a high, $\approx20\times10^{4}\,\textrm{km}\,\textrm{s}^{-1}$, velocity, corresponding to $\rho t^3\approx10^{-3}M_\odot/(10^4{\rm km\,s^{-1}})^3$. To demonstrate the effect of such a fast component, we added to our $1\,M_{\odot}$ model a high velocity component of $^{56}$Co, around $\rho t^3=10^{-3}M_\odot/(10^4{\rm km\,s^{-1}})^3$, with $10\%$ of the mass of the original model. The $\gamma$-rays from the fast component escape at early times ($t_0\approx13\,\textrm{day}$), so that the $^{56}$Ni mass deduced from the light curve would not include the fast component. We therefore plot in Figure~\ref{fig:deposition} the calculated $Q_d(t)$ for this model as compared to $Q_{ins}(t)$ without the fast component. While the mass contained in the fast component is small, it increases the heating rate by a factor $\mysim2$ at late times. This effect may thus be used to constrain the velocity distribution of the synthesized $^{56}$Ni.

\subsection{The validity of the $L=Q$ approximation}
\label{sec:LQ}

We have approximated the bolometric luminosity as given by the energy deposition rate, $L=Q$. This is a valid approximation provided that both the plasma cooling time, $t_{\rm{cool}}$, and the (optical) photon escape time, approximately given at the late time of interest by the light crossing time of the ejecta $r/c$, are shorter than the characteristic time for changes in the deposition rate, $\max[t_{1/2},\tau_{\rm dep}]$ (we ignore here adiabatic energy losses of the expanding heated plasma, which may become significant at $t>t_\varepsilon$, i.e. when $\tau_{\rm dep}>t$, since we are interested mainly in the time range of $t_\tau<t<t_\varepsilon$, due to the fact that at $t>t_{\varepsilon}$ the delayed deposition contribution to the heating of the plasma is dominated by the decays of $^{57}$Co and $^{55}$Fe). The latter condition is satisfied since at $t=t_\tau$ we have $r/c=(v/c)t_\tau\approx20\,{\rm d}<t_{1/2}$ (using Eq.~(\ref{eq:t tau}) and $v/c=1/30$), and at later time
$\tau_{\rm dep}\propto\rho^{-1}\propto t^3$. Let us consider next the former condition. As long as $t_{\rm{cool}}\ll \max[t_{1/2},\tau_{\rm dep}]$, the cooling rate equals the heating rate and is approximately given by $t_{cool}\approx1\textrm{eV}/q$, where $q$ is the energy deposition rate per atom \citep{Axelrod1980}. Using the instantaneous deposition value, i.e. neglecting the delayed deposition effect, of $q$ for $^{56}$Co $\beta^{+}$ positrons, given by Equation~\eqref{eq:56 deposition}, we have $t_{cool}\approx73\exp(t/111.4\,\textrm{d})\,\textrm{s}$ and $t_{\rm{cool}}< t_{1/2}$ at $t\lesssim1300\,\textrm{day}$. This implies that in fact the cooling time is much smaller than $\max[t_{1/2},\tau_{\rm dep}]$ at $t=1300$~day, since at this time $t_{1/2}\ll\tau_{\rm dep}$ and $q$ is larger than the instantaneous value given by Equation~\eqref{eq:56 deposition} due to the delayed deposition. The approximation of $L=Q$ holds therefore well beyond 1300~day, typically to $\gtrsim2000$~day. 

A comment is in place here also regarding the recombination energy. Since the plasma cannot be assumed to be in an ionization equilibrium at $t\gtrsim600\,\textrm{day}$ \citep{Axelrod1980}, the recombination energy fraction of the deposited energy may be stored in the plasma without being radiated away (reducing $L$ below $Q$). However, the recombination energy is typically only a few percent of the effective average ionization energy, $\bar{I}$ (see Equation~\eqref{eq:ionization losses} and the text below). Most of the deposited energy is therefore transferred to the plasma by the kinetic energy loss of the secondary electrons, and the correction to our $L=Q$ approximation is small.


\section{The case of SN 2011\lowercase{fe}}
\label{sec:SN 2011fe}

\begin{figure}
\includegraphics[width=0.48\textwidth]{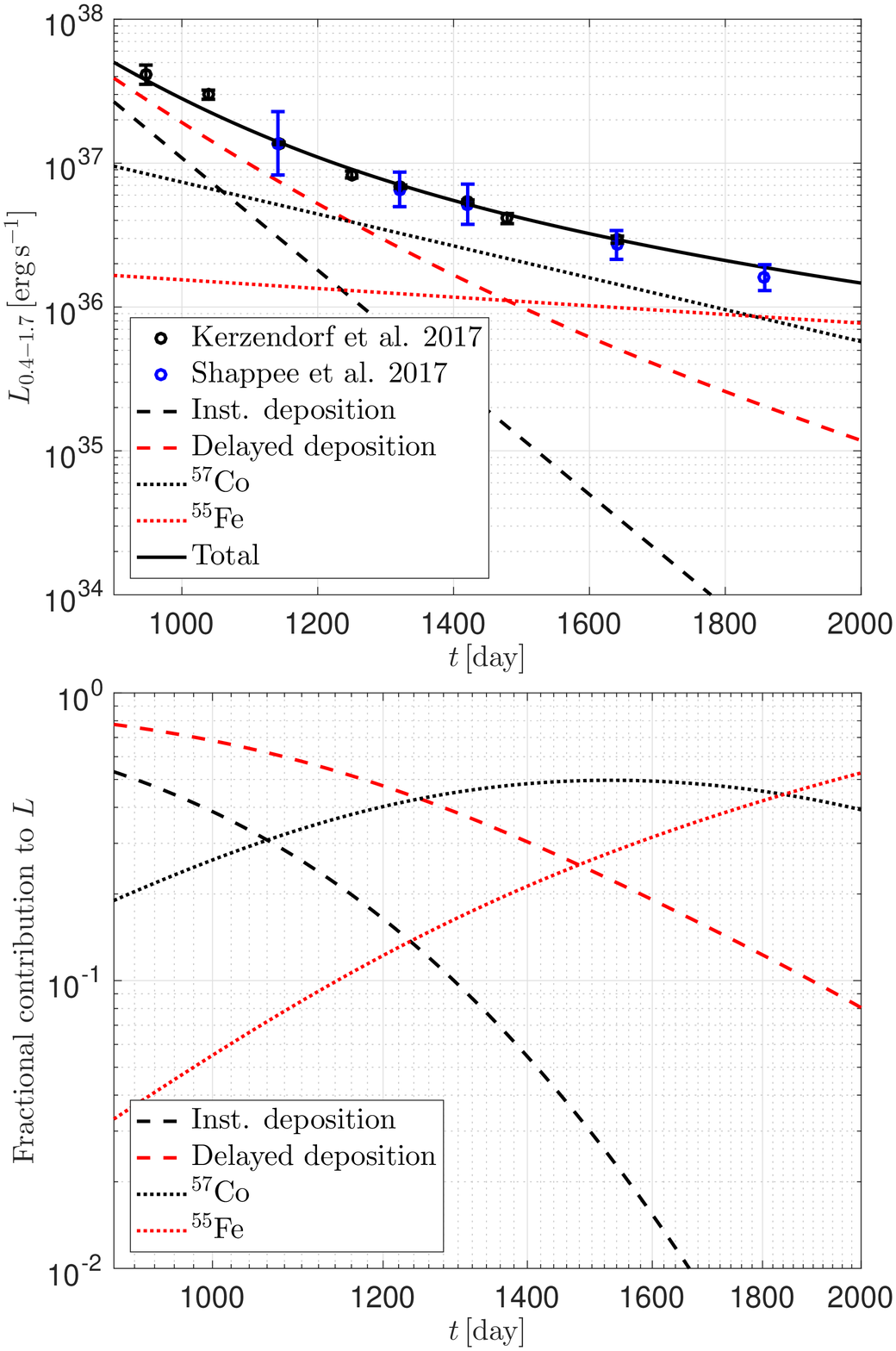}
\caption{The best-fitting model to the late-time quasi-bolometric light curve, $L_{0.4-1.7}$, of SN 2011fe. Top panel: black circles -- the data of \citet{Kerzendorf2017}, blue circles -- the data of \citet{Shappee2017}. We fit for $\langle \rho  \rangle t^3$ and for the fraction of observed flux (compared to the total bolometric emission), under the following assumptions: total $^{56}$Ni mass of $0.5\,M_{\odot}$, a distance of $6.4\,\textrm{Mpc}$, a singly ionized plasma ($\chi_e=1$) with a solar metallicity (determining the $^{57}$Co$/^{56}$Co and $^{55}$Fe$/^{56}$Co mass ratios), and the time of maximum flux in the $B$ band corresponding to $18\,\textrm{day}$ after the explosion \citep{Pereira2013}. The instantaneous deposition (black dashed line) is shown only for comparison. We obtain $\langle \rho t^3 \rangle_{\rm{SND}}\approx0.14$, consistent with the sub-Chandra and direct-collision models, and observed flux fraction of $\approx0.37$. Bottom panel: The fractional contribution of each component to $L$, with the same lines as in the top panel (here too, the instantaneous deposition is shown only for comparison). The delayed deposition (red dashed line) is significantly larger than the instantaneous deposition and is the dominant contribution to the light curve for $t\lesssim1200\,\textrm{day}$. At later times the light curve is dominated by $^{57}$Co (dotted black line) and by $^{55}$Fe (dotted red line).
\label{fig:sn2011fe}}
\end{figure}

The best Type Ia supernova for late-time studies is SN 2011fe \citep[see discussion in][]{Shappee2017}, which was observed over a wide range of wavelengths at late time. However, even in this case, a significant fraction of the flux is predicted to be emitted outside of the observed wavelength range \citep{Fransson2015}, complicating the comparison to models. For this supernova, the bolometric correction is more uncertain at $\sim600-800\,\textrm{days}$ \citep{Dimitriadis2017}, where NIR data are not available. We therefore use only the $t\gtrsim900\,\textrm{day}$ data \citep[black and blue circles in Figure~\ref{fig:sn2011fe} showing the quasi-bolometric, $4000-17000$ \AA, light curve,][respectively]{Kerzendorf2017,Shappee2017}. At these times, the $\gamma$-ray contribution to the light curve may be neglected (see Figure~\ref{fig:deposition}).

We do not provide a formal fit to the light curve, but instead demonstrate the delayed deposition effect with a best-fitting model under some simplifying assumptions. This is due to the fact that in order to fit a model to the quasi-bolometric light curve, we make the assumption that the fraction of unobserved flux is constant within the time span of the fit. Since this fraction is significant, this assumption may lead to a significant systematic error. We also note that in the four epochs for which both \citet{Shappee2017} and \citet{Kerzendorf2017} provide quasi-bolometric fluxes, the ratio between the error bars of two compilations is $\sim4-30$.

We fit for $\langle \rho  \rangle t^3$ and for the fraction of observed flux, under the following assumptions: total $^{56}$Ni mass of $0.5\,M_{\odot}$, a distance of $6.4\,\textrm{Mpc}$, a singly ionized plasma ($\chi_e=1$) with a solar metallicity (determining the $^{57}$Co$/^{56}$Co and $^{55}$Fe$/^{56}$Co mass ratios), and the time of maximum flux in the $B$ band corresponding to $18\,\textrm{day}$ after the explosion \citep{Pereira2013}. The best-fitting solution, which fits the data well, is shown in Figure~\ref{fig:sn2011fe}. We obtain $\langle \rho t^3 \rangle_{\rm{SND}}\approx0.14$, consistent with the sub-Chandra and direct-collision models, and an observed flux fraction of $\approx0.37$. The delayed deposition (red dashed line) is significantly larger than the instantaneous deposition (black dashed line) and is the dominant contribution to the light curve for $t\lesssim1200\,\textrm{day}$. At later times the heating and the bolometric light curve are dominated by $^{57}$Co (dotted black line) and by $^{55}$Fe (dotted red line) decays. The effect of the delayed deposition is to decrease the inferred amount of $^{57}$Co and $^{55}$Fe. However, a reliable determination of the parameters would require a more complete measurement of the bolometric light curve.


\section{Summary and discussion}
\label{sec:discussion}

We have calculated the effect of the finite time, $\tau_{\rm dep}$, given by Equation~(\ref{eq:tau_dep}), over which positrons from $\beta^{+}$ decays of $^{56}$Co deposit their energy in the ejecta of type Ia supernovae on the late-time light curves of these explosions, assuming that the positrons are trapped within the ejecta. The equations describing the deposition of energy are given in Sections~\ref{sec:e energy} and~\ref{sec:e_dep}. A Matlab code that solves these equations, and is straightforward to use for the analysis of observed light curves, is provided (see footnote 1).

We have shown that a significant light-curve flattening is obtained due to delayed deposition at $t\gtrsim t_\tau$, where $t_\tau$, given by Equation~(\ref{eq:t tau}), is the time at which the ejecta density drops below the value for which $\tau_{\rm dep}$ equals the $^{56}$Co life time, $t_{1/2}$. We find that the ejecta heating is dominated by delayed deposition typically from $600$ to $1200\,\textrm{day}$, and only later by the decay of the longer lived isotopes $^{57}$Co and $^{55}$Fe (assuming solar abundance), see Figures~\ref{fig:deposition} and~\ref{fig:sn2011fe}.

The modification of the light curve is determined by the density distribution of the ejecta. We have shown (see Figure~\ref{fig:deposition}) that for the relatively narrow velocity distributions of commonly studied explosion models, including Chandra, sub-Chandra, and direct-collision models, the effect depends mainly on the $^{56}$Ni mass-weighted average density $\langle \rho \rangle t^{3}$. Nevertheless, the strong dependence of $t_\tau$ on the velocity implies that the presence of a relatively small amount of $^{56}$Co within fluid elements with relatively high velocity, with correspondingly low values of $\rho t^{3}$, may have a large effect on the delayed deposition rate and bolometric luminosity, see Figure~\ref{fig:deposition}. This is due to the fact that the time at which delayed deposition would become important for such fast components would be significantly earlier than $t_\tau$ corresponding to the average density.

Accurate late-time bolometric light curves, which may be obtained with far-IR measurements as would be provided by JWST, will thus enable one to discriminate between explosion models by determining $\langle \rho \rangle t^3$ and the $^{57}$Co and $^{55}$Fe abundances, and by constraining the width of the velocity distribution of the synthesized $^{56}$Ni.

In Section~\ref{sec:SN 2011fe}, we have analysed the late-time light-curve observations of SN 2011fe, for which the most complete late-time wavelength coverage is available. We have shown that the observed flattening is consistent with $\langle \rho\rangle t^{3} \approx 0.15\,M_{\odot}\,(10^{4}\, \textrm{km}\,\textrm{s}^{-1})^{-3}$ and solar abundance $^{57}$Co and $^{55}$Fe (see Figure~\ref{fig:sn2011fe}). In general, the contribution of delayed deposition to the heating of the ejecta at $t>600\,\textrm{day}$ implies that the recent observations suggesting late-time light-curve flattening are readily accounted for by delayed deposition, and do not imply supersolar $^{57}$Co and $^{55}$Fe abundances.

While the delayed deposition effect is important for Ia SNe, it is unlikely to be observed in CCSNe, due to their larger values of $\langle \rho  \rangle t^{3}$, as shown in Figure~\ref{fig:rhot3}. The larger densities imply larger values of $t_\tau$, so that by the time delayed deposition of $^{56}$Co positrons becomes important, the energy deposition in the ejecta is already dominated by the decay of the longer lived isotopes $^{57}$Co and $^{55}$Fe, see Figure~\ref{fig:to_t_demo} \citep[for a solar metallicity mass ratio, $^{57}\textrm{Fe}/^{56}\textrm{Fe}=0.023$,][the instantaneous energy deposition rates of $^{56}$Co and $^{57}$Co are equal at $t_{57}\approx1050\,\textrm{day}$]{Lodders2019}. We note that the energy deposition by $^{57}$Co and $^{55}$Fe may be considered as instantaneous, since the energy of the electrons produced is lower, with energy loss rate significantly higher, compared to that of $^{56}$Co positrons, and since the life times of $^{57}$Co and $^{55}$Fe are significantly longer than that of $^{56}$Co. The combination of the shorter loss time and longer decay time implies that, at the relevant times, the energy deposition time is small compared to the $^{57}$Co and $^{55}$Fe life times, and the effect of delayed energy deposition is small.

We note in this context that the escape of positrons, in case they are not fully trapped, may affect the light curves of CCSNe only at very late time. In the extreme case of the absence of magnetic fields, the ejecta crossing time of the escaping positrons is $t_{\rm{cross}}\sim vt/v_e$, where $v\sim10^{4}\,\textrm{km}\,\textrm{s}^{-1}$ is the typical velocity of supernovae ejecta. The suppression of energy deposition due to escape is significant only if positions escape before losing their energy to heating, i.e. only if $t_{\rm{cross}}<\tau_{\rm{dep}}$. Using Equation~\eqref{eq:t eps}, we find that this condition holds at $t>t_{\rm{esc}}\approx(v/v_e)^{1/2}t_{\varepsilon}$, where $t_{\rm{esc}}$ is the time at which $\tau_{\rm{dep}}=t_{\rm{cross}}$. For the densities relevant for Type Ib/c SNe, $\langle \rho t^3 \rangle_{\rm{SND}}\approx1\,{\rm to\,} 10$ (see Figure~\ref{fig:rhot3}), $t_{\rm{esc}}\gtrsim700\,\textrm{day}$, so that considering positron escape at much earlier times for these SNe \citep[e.g.][]{Wheeler2015} is unrealistic. For SNe of type II the densities are expected to be higher, and $t_{\rm{esc}}$ is expected to be still larger.

Several comments are in place here regarding the differences between the Ia SN case discussed in this paper and the kilonova case discussed by \citet{Waxman2018,Waxman2019}. First, the characteristic density of the ejecta is some four orders of magnitude lower in the kilonave case, due to the smaller ejecta mass and its higher velocity. Second, in the kilonova case the radioactive energy release is dominated at early times $t$ by the decay of isotopes with $t_{1/2}\sim t$ so that $t_{\tau}$ is similar to the time $t_{\varepsilon}$, given by Equation~(\ref{eq:t eps}), at which adiabatic losses of the electrons/positrons become significant (for Ia SNe, the two times are distinct, with $t_{\tau}<t_{\varepsilon}$). As a result, the characteristic time-scales in the kilonva case are $\sim10\,\textrm{day}$, instead of $\sim1000\,\textrm{day}$ in the Ia SN case. Finally, the velocity distribution of radioactive material is wide in the kilonova case, and the light-curve modifications depend not only on the average density but also on the shape of the density distribution. In particular, due to the strong dependence of $t_{\tau} \sim t_{\varepsilon}$ on the velocity, the presence of a component with small mass at high velocity may strongly affect the light curve at times earlier than those estimated using $\langle \rho \rangle t^3$. The behavior of the kilonovae light curves at $t>t_{\varepsilon}$ was analysed in detail by \citet{Waxman2019}. For the Ia SNe discussed here, where $t_\tau<t_{\varepsilon}$, we were interested mainly in the regime $t_\tau<t<t_{\varepsilon}$ since at $t>t_{\varepsilon}$ the contribution to the heating of the plasma is dominated by the decays of $^{57}$Co and $^{55}$Fe.

\section*{Acknowledgements}
We thank Boaz Katz for useful discussions. We thank Luc Dessart, Sung-Chul Yoon, Tuguldur Sukhbold, Sergei Blinnikov, and Victor P. Utrobin for sharing their supernovae models with us. DK is supported by the Israel Atomic Energy Commission -- The Council for Higher Education -- Pazi Foundation -- and by a research grant from The Abramson Family Center for Young Scientists. EW is partially supported by ISF, GIF, and IMOS grants.






\bsp	
\label{lastpage}
\end{document}